\documentstyle[aps,prb]{revtex} \def\narrowtext{} \tighten \twocolumn
\begin{document}
\draft
\title{ARPES Study of the Superconducting Gap Anisotropy in 
$Bi_2Sr_2CaCu_2O_{8+x}$}
\author{
        H. Ding$^{1,2}$,
        M. R. Norman$^{1}$,
        J. C. Campuzano$^{1,2}$,
        M. Randeria$^{3}$,
        A. F. Bellman$^{1,4}$,
        T. Yokoya$^5$, T. Takahashi$^5$,
        T. Mochiku$^6$, and K. Kadowaki$^6$
       }
\address{(1) Materials Sciences Division, Argonne National Laboratory,
             Argonne, IL 60439 \\
         (2) Department of Physics, University of Illinois at Chicago,
             Chicago, IL 60607\\
         (3) Tata Institute of Fundamental Research, Bombay 400005, India\\ 
         (4) Dipartimento di Fisica, Universita di Milano, 20133 Milano, 
             Italy,\\
         (5) Department of Physics, Tohoku University, 980 Sendai, Japan\\
         (6) National Research Institute for Metals, Sengen, Tsukuba,
             Ibaraki 305, Japan\\
         }

\address{%
\begin{minipage}[t]{6.0in}
\begin{abstract}
We report measurements of the momentum dependence of the 
superconducting gap in $Bi_2Sr_2CaCu_2O_{8+x}$ (Bi2212)
with angle-resolved photoemission 
spectroscopy using a dense sampling of the Brillouin
zone in the vicinity of the Fermi surface.  In the $Y$ quadrant of the zone,
where there are no complications from ghost bands caused by the
superlattice, we find a gap function consistent within error
bars to the form $\cos(k_x)-\cos(k_y)$ expected for a d-wave order parameter.
Similar results are found in the $X$ quadrant
with the photon polarization chosen to enhance main band
emission over that due to ghost bands.
\typeout{polish abstract}
\end{abstract}
\end{minipage}}

\maketitle
\narrowtext

The past few years have seen an intense effort on the part of the condensed
matter physics community to determine the form of the order parameter
in the high temperature cuprate superconductors \cite{review}.
Part of this effort has relied on the use of 
high resolution angle-resolved photoemission spectroscopy (ARPES) 
since this is the only known technique
which can directly map out the momentum dependence of the superconducting gap 
\cite{olson,shen,kelly,ding95}.

Very recently there has been considerable progress
in ARPES studies of the cuprates. This includes
evidence for a spectral function interpretation of ARPES
via the study of sum rules \cite{randeria95}, a detailed
study of the Fermi surface \cite{ding96} and
momentum distribution \cite{campuzano96} in Bi2212, 
and an elucidation of the polarization selection rules 
\cite{ding96,norman95b} in the presence
of the structural superlattice, which necessarily affects ARPES data
on the highest quality Bi2212 samples.
All these points are crucial in accurately
determining the momentum dependence of the gap.

Modeling the experimental data in terms of spectral functions
gives a quantitative estimate of the gap \cite{ding95}
at a given location on the Fermi
surface, but perhaps even more importantly, 
it provides a framework within which one can think about the data.
This points to the importance of having a very fine momentum
space grid for data, and also 
for a very precise determination of the Fermi surface,
in the neighborhood of the region where nodes in the gap
occur. Further, the complications \cite{norman95b}
arising from the superlattice
make it simplest to study the gap anisotropy in the
$Y$ quadrant \cite{notation}, 
since here the main CuO Fermi surface is widely separated from
the ``ghost'' Fermi surfaces due to the superstructure \cite{superlattice}.

In this paper, we summarize the results obtained on several samples
of Bi2212 with experiments which were designed 
taking into account all of the points listed above.
We find that, within error bars, the excitation gap on the 
main Fermi surface sheet is consistent with 
$\vert \cos(k_x) - \cos(k_y) \vert$. The ARPES experiment cannot
of course measure the phase of the order parameter, but this result
is strongly suggestive of a $d_{x^2-y^2}$ order parameter. 
Finally, in order to prove that the two node gap previously
seen in the $X$ quadrant \cite{ding95} 
was indeed an artifact \cite{norman95b} of the superstructure,
we also study the $X$ quadrant
using a polarization geometry which enhances the main band relative to the
``ghost'' band, and obtain results which are in agreement with the $Y$ quadrant.

The experiments were done at the Synchrotron Radiation Center,
Wisconsin, using a high resolution 4-m normal incidence monochromator with
a resolving power of $10^4$ at $10^{11}$ photons/s.  
We used 22 eV photons, with a 17 meV (FWHM) energy resolution, and a
momentum window of radius 0.045$\pi$ 
(in units \cite{notation} of $1/a^*$).  

The crystals, which were grown by the 
traveling solvent floating zone method with an infrared 
mirror furnace, have low defect densities with
structural coherence lengths $\sim 1250\AA$ obtained from X-ray 
diffraction. The samples were cleaved in-situ at 13 K in a vacuum of
$<$ 5x$10^{-11}$ Torr.  
Most samples have very flat surfaces after cleaving,
as measured by specular laser reflections.  A flat surface is crucial for
this experiment since it directly affects the momentum resolution.
Another measure of
the sample quality is the observation of the ``ghost" bands due to the
superlattice distortion; in our best sample (87K $T_c$,
with a 1K transition width) we have now also seen
evidence for the second umklapps from the superlattice.
We will also present data from three other
samples with a 90K $T_c$.

In Fig.~1, we show the 13K experimental energy distribution curves (EDCs)
for the 87K $T_c$ sample for various locations on the main 
band Fermi surface (FS) in the $Y$ quadrant. 
The spectra shown correspond to the minimum observable
gap along a set of ${\bf k}$ points normal to the FS (for a detailed
discussion of this in the context of a study of particle-hole mixing,
see Ref.~\onlinecite{campuzano96}).
These spectra are obtained from a dense sampling of ${\bf k}$-space
in the vicinity of the FS which is almost five times denser
than previous data.  In Ref.~\onlinecite{ding95} 
a ${\bf k}$-step size of 0.064$\pi$ normal 
to the FS and 0.064-0.081$\pi$ along the FS
was used in the $Y$ quadrant (the radius of the ${\bf k}$-window is 0.045$\pi$).
The problem is that the bands are highly dispersive along $\Gamma Y$ with an
energy change of about 85 meV per 0.064$\pi$ step, and thus ${\bf k}_F$
cannot be located accurately.  Therefore, in our
new measurements, we use a step size of 0.0225$\pi$ normal to the Fermi
surface and 0.045$\pi$ along the Fermi surface.
This not only allows us to map out the nodal
region more clearly, but also improves by a factor of about three our
ability to locate that spectrum whose binding energy 
at the center of the momentum
window is closest to the Fermi energy (that is, the step size normal to
the Fermi surface now corresponds to a dispersion of about 30 meV per step).  
In addition, at each ${\bf k}$ point 
the photon polarization is chosen along $\Gamma X$ so as to maximize
emission on the $\Gamma Y$ diagonal direction, i.e., $\Gamma Y\perp$ geometry
(a polarization rotated $45^\circ$ relative to this 
was used in Ref.~\onlinecite{ding95}).

In each panel of Fig.~1 we also plot the spectrum of a
platinum reference, in electrical contact with Bi2212, 
measured periodically to determine the chemical potential, 
and to check for drifts in beam energy. 
(The polycrystalline Pt spectrum is a weighted density of states
whose leading edge is an energy-resolution limited 
Fermi function).  The simplest gap estimate is
obtained from the mid-point shift of the leading edge of Bi2212 
relative to Pt.  This has no quantitative validity 
(since the Bi2212 EDC is a spectral
function \cite{randeria95}, while Pt is a density of states)
but yields an angular dependence which is qualitatively similar
to the results described below.

The simplest way to quantitatively estimate the gap
is to model \cite{ding95}
the data in terms of a simple BCS lineshape formula,
taking into account the measured energy dispersion
and the known energy and momentum resolutions. 
Two important points need to be discussed in connection 
with such fits: first, the lack of knowledge about the 
spectral lineshape, especially its incoherent part,
and, second, the large background in Bi2212
whose origin is unclear. In the large gap region near the
$\bar{M}$ point, we see a linewidth (imaginary self-energy) collapse, 
for frequencies smaller than 3$\Delta$,
upon cooling well below $T_c$ \cite{randeria95,ding96}. Thus the
coherent piece of the spectral function is modeled by
the BCS lineshape, with all of the incoherent part lumped together
with the experimental background in so far as the fitting procedure 
is concerned. We also showed \cite{ding95} that it was self-consistent
(though perhaps not unique) to make the same assumption in the
small gap region; the much larger width of the EDC in the diagonal
direction arising due to the ${\bf k}$ resolution combined
with the large dispersion.

The gaps extracted by fitting the spectra of Fig.~1 are shown in Fig.~2.
We emphasize that since the gap is determined by fitting the 
resolution-limited leading edge of the EDC, it is relatively unaffected 
both by self-energy effects, and by the experimental background 
which cuts off at low frequencies. To check this, we have made an
independent set of fits to the small gap data
where we do not use any background fitting function,
and only try to match the leading edges, not the full spectrum.
The two gap estimates are consistent within a meV.
The (vertical) error bars of $\pm$ 3meV on the gaps in Fig.~2 
come primarily from 
the quality of the fit to the leading edge of the 
data and uncertainty in the location of the center of
the ${\bf k}$-window (near the diagonal direction), 
with smaller contributions coming from 
chemical potential determination
($\pm$ 0.5meV) and background modeling ($\pm$ 0.5meV).  
Horizontal error bars represent the accuracy in
which we can determine the Fermi surface angle.  
This does not include any
effective error bar coming from ${\bf k}$ resolution, 
since this is in principle taken into account in the fits.
The angular variation 
of the gap in Fig.~2 is in excellent agreement
with that expected from a d-wave order parameter of the 
$\cos(k_x) - \cos(k_y)$ form\cite{yquad_fit}.

Next we turn to the $X$ quadrant of the 87K $T_c$ sample.
It is now recognized \cite{norman95b} that the two node gap observed
previously \cite{ding95} in the $\Gamma X \vert\vert$ geometry
came from an even linear combination of the ``ghost'' bands
due to the superlattice \cite{superlattice}.
We thus choose the polarization along $\Gamma Y$, so that on the diagonal
we are in a $\Gamma X\perp$ geometry which enhances emission from the
main band \cite{odd_sl}. 
The $X$ quadrant gaps, determined from spectral function fits,
are plotted in Fig.~3.
We see that the hump in the gap along $\Gamma X$ ($45^\circ$)
seen previously \cite{ding95} has indeed disappeared.
The solid curve is a fit of the data to a d-wave gap function
with a small sample misalignment ($1.4^\circ$ in real space).
Note that for this data set, the step size along the Fermi surface was
0.135$\pi$ and so is not dense enough around $45^\circ$ 
to address the question of the detailed behavior around the node.

We next summarize the results (Fig.~4)
for $Y$ quadrant gaps extracted from fits
on three different 90K $T_c$ samples.
The main point is to note possible complications
which arise in interpreting data sets which are not as
dense in ${\bf k}$ space as 
the detailed $Y$ quadrant measurements on the 87K sample described above
(the step size along the Fermi surface being twice as large).
The results on sample I have a region of reduced gap, consistent with
zero, near 45$^\circ$.  
To some extent this may be an artifact of
the finite diameter of the ${\bf k}$-window, which is
6$^\circ$ in FS angle \cite{gap_k}. 
In addition, we found that for the small gap
points of sample I the leading edge of the data 
lies above, i.e., to the right of,
that of a zero gap spectral fit, assuming the ${\bf k}$-window 
was centered at ${\bf k}_F$. 
We find that a combination of factors (${\bf k}$-window center, 
chemical potential drift, and background) discussed above can indeed
conspire to produce such an anomalous shift. 
These factors are already reflected in the $\pm 3$ meV
error bars on the gap estimate, but we reiterate that
the error bars must be borne in mind while
interpreting the final results.

Similar issues arise with the results on sample II, and in addition both this
and sample III appear to have an asymmetry about the diagonal (45$^\circ$).
For sample III, the asymmetry is quite large and almost certainly due to
sample misalignment (of $1.8^\circ$ in real space);
the sample III results in Fig.~4 have been
shifted by the corresponding FS angle to facilitate comparison 
with the other two samples.
All of these caveats would also apply to the sparse $Y$ quadrant data
presented in Ref.~\onlinecite{ding95}, and, in particular, 
the shift of the node away from 45$^\circ$ in that data 
may also have been a consequence of sample misalignment.  

In summary, we would say that the results
of the 90K samples are consistent,
within error bars, with those obtained from
the detailed measurements on the 87K sample.
There is no hard evidence for either
an extended region of nodes about 45$^\circ$,
or for a mixed symmetry gap.

In conclusion, we have determined the intrinsic momentum dependence of the
superconducting gap in Bi2212 using dense momentum sampling of 
the zone in the vicinity of the Fermi surface.  
We find that the observed gap is consistent with the
simple d-wave form \cite{d_wave} $\cos(k_x)-\cos(k_y)$ 
within experimental error bars.

This work was supported by the U.~S.~Department of Energy,
Basic Energy Sciences, under Contract \#W-31-109-ENG-38 and the National
Science Foundation (DMR 91-20000) through the Science and Technology Center
for Superconductivity.  The Synchrotron
Radiation Center is supported by NSF grant DMR-9212658.

\begin{figure}
\caption{Bi2212 spectra (solid lines) for a 87K $T_c$ sample at 13K
and Pt spectra (dashed lines)
versus binding energy (meV) along the Fermi surface in the $Y$ quadrant,
with locations shown in Fig.~2.}
\label{fig1}
\end{figure}

\begin{figure}
\caption{$Y$ quadrant gap in meV versus angle on the Fermi surface
(filled circles) with fits to the data using a d-wave gap (solid curve).
Labels of data points correspond to the spectra of Fig.~1.  Inset shows
their locations in the zone as well as the photon polarization direction.}
\label{fig2}
\end{figure}

\begin{figure}
\caption{$X$ quadrant gap in meV, for the 87K $T_c$ sample,
measured at 13K, versus angle on the Fermi surface
(filled circles) with fits to the data using a d-wave gap (solid curve).
The photon polarization is along $\Gamma Y$.}
\label{fig3}
\end{figure}

\begin{figure}
\caption{$Y$ quadrant gap in meV, measured at 13K, 
versus angle on the Fermi surface
for three different Bi2212 samples each with a 90K $T_c$.  
For visual clarity only a representative error bar has been shown.}
\label{fig4}
\end{figure}


\begin{references}

\bibitem{review} 
B. Goss-Levi, Physics Today, {\bf 49}, 1, p.~19 (1996). 

\bibitem{olson}
C.G. Olson et al., Science {\bf 245}, 731 (1989).

\bibitem{shen} 
Z.-X. Shen et al., Phys. Rev. Lett. {\bf 70}, 1553 (1993).

\bibitem{kelly} R. J. Kelley et al., Phys. Rev. B {\bf 50}, 590 (1994).

\bibitem{ding95} H. Ding et al., Phys. Rev. Lett. {\bf 74}, 2784 (1995)
and {\bf 75}, 1425 (E) (1995).

\bibitem{randeria95}
M. Randeria et al., Phys. Rev. Lett. {\bf 74}, 4951 (1995).

\bibitem{ding96} H. Ding et al., Phys. Rev. Lett. {\bf 76}, 1533 (1996).

\bibitem{campuzano96} J. C. Campuzano et al., preprint, cond-mat/9602119.

\bibitem{norman95b} M. R. Norman et al., Phys. Rev. B {\bf 52}, 15107 (1995).

\bibitem{notation}
We use a square lattice notation with the axes along the
CuO bond directions. $X=(\pi,-\pi)$ and $Y=(\pi,\pi)$
in units of $1/a^*$, where $a^* = 3.83\AA$ is the separation 
between near neighbor Cu ions. 
Thus the orthorhombic a axis is along $X$ and b axis along $Y$.

\bibitem{superlattice}
The incommensurate superlattice in Bi2212 causes
images (``ghost'' bands) of the main CuO
band. The first umklapps, always seen in good samples,
are displaced from the main band by $\pm$ the superlattice vector
${\bf Q} =  (0.21\pi,0.21\pi)$.
In the $Y$ quadrant, the bands and the corresponding Fermi surfaces
are widely separated, since ${\bf Q}$ is along $\Gamma Y$; however
this is not so in the $X$ quadrant; 
see Ref.~\onlinecite{ding96} and
M. R. Norman et al., Phys. Rev. B {\bf 52}, 615 (1995).

\bibitem{yquad_fit}
The actual functional form is $\cos(2\phi+\phi_0)$, 
where $\phi_0$, with a best fit value $1.4^\circ$
(in FS angle), represents a possible sample misalignment 
(in real space) of $\simeq 0.4^\circ$.
This form agrees very closely with $\cos(k_x)-\cos(k_y)$ over the observed
FS up to a multiplicative constant of 0.46.

\bibitem{odd_sl}
In principle an odd ``ghost'' band, with the same
symmetry as the main band, should also contribute in this geometry
\cite{norman95b}. There is, however, no evidence \cite{ding96}
in the data for such an odd ``ghost'' band, possibly due to final 
state effects. We note that the emission intensities from 
the main and ghost bands are not well understood at this time.

\bibitem{gap_k} So as not to prejudice the fits, the gap was considered to be
a constant within the momentum window.  This condition is obviously violated
in the d-wave case near the node where the gap is varying rapidly with momentum.

\bibitem{d_wave} Such an order parameter arises naturally
from theories with antiferromagnetic spin fluctuations and/or
strong correlations.  For recent reviews, see D. Pines and P. Monthoux,
J. Phys. Chem. Solids {\bf 56}, 1651 (1995) and
D. J. Scalapino, Phys. Rep. {\bf 250}, 329 (1995) and J. Phys. Chem. Solids
{\bf 56}, 1669 (1995).

\end{references}
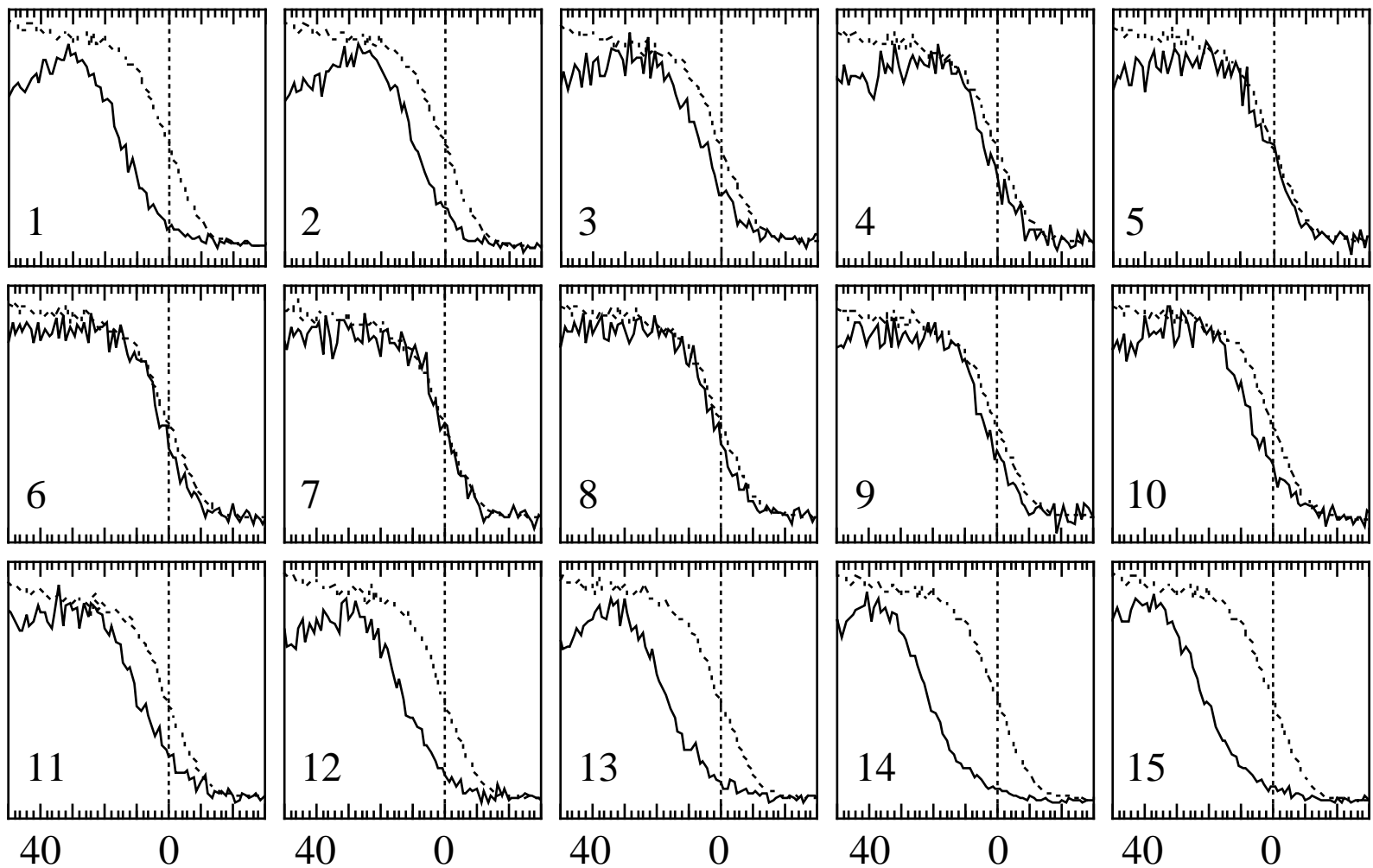
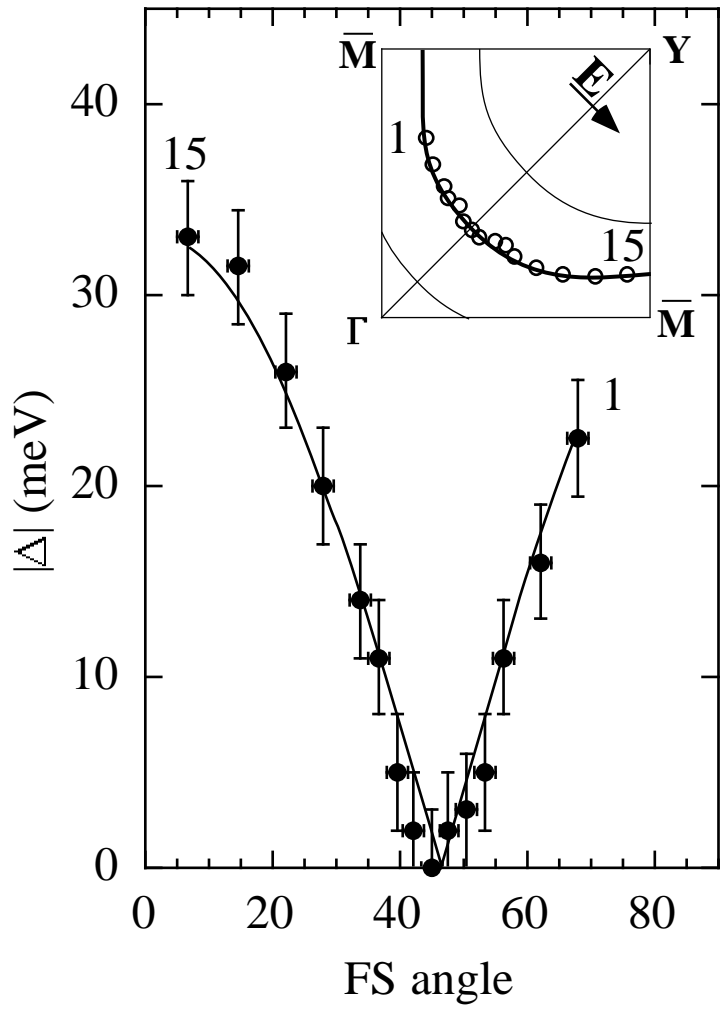
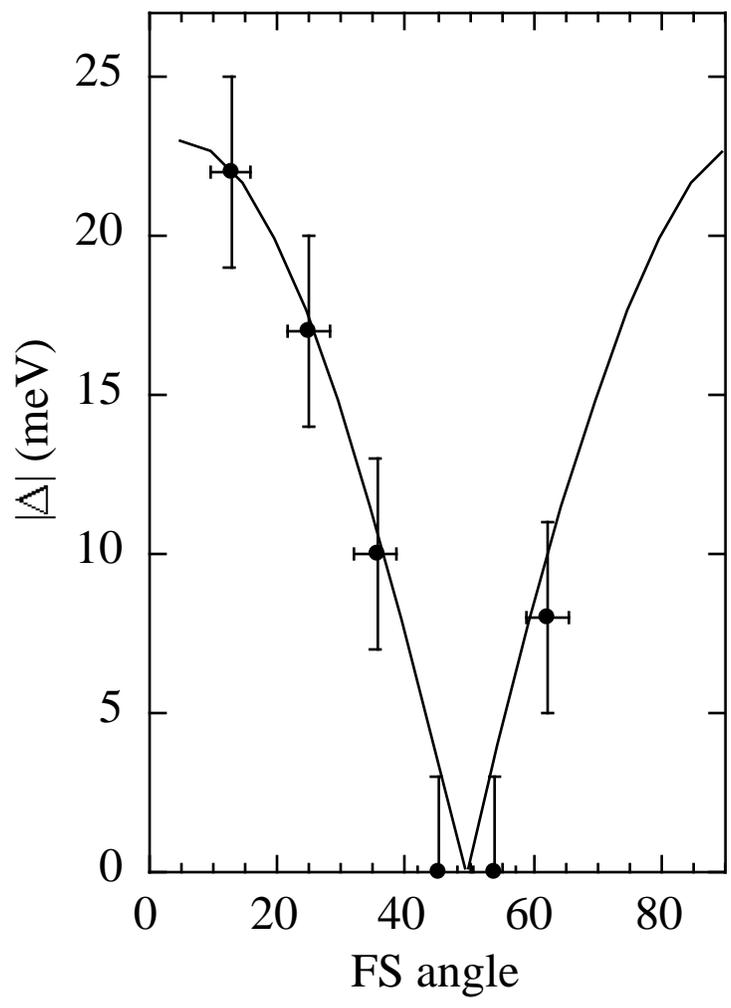
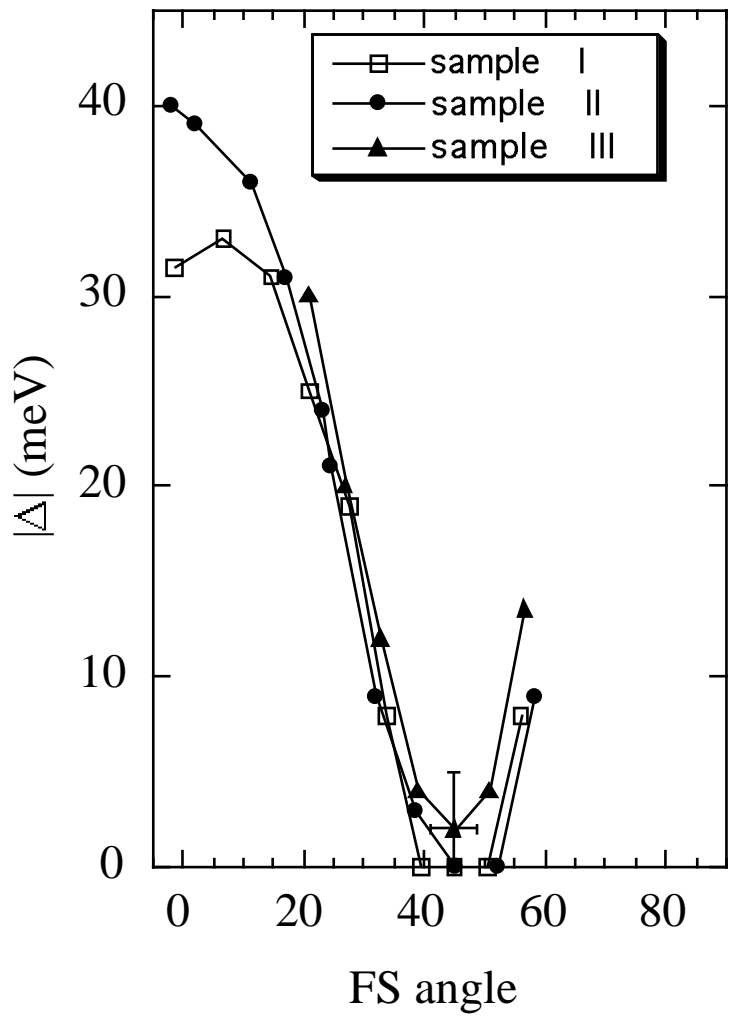
\end{document}